\documentclass[conference]{IEEEtran}
\usepackage{cite}
\usepackage{url}
\usepackage{balance}
\usepackage{amsmath,amssymb,amsfonts}
\usepackage{algorithmic}
\usepackage{graphicx}
\usepackage{textcomp}
\usepackage{xcolor}
\usepackage{comment}
\newcommand{\spacetune}[1]{#1}

\begin{document}

\title{
\mbox{A Computational Toolkit for}
\mbox{Engagement and Scalable Assessment}
\mbox{in a Large Logic Course}
}

\author{%
\IEEEauthorblockN{%
Stephen M. Watt}
\IEEEauthorblockA{\newline\textit{Cheriton School of Computer Science} \\
\textit{University of Waterloo}\\
Waterloo, Ontario, Canada \\
smwatt@uwaterloo.ca}
}

\maketitle

\begin{abstract}
Large required courses in theoretical computer science face two related
challenges: helping students engage with abstract material and supporting
reliable student assessment at scale.  This paper describes Logicat, a
lightweight computational toolkit developed for CS 245, \textit{Logic and
Computation}, at the University of Waterloo.  The course is required for
undergraduate computer science students and serves a large annual cohort.

The main pedagogical objective is to help students concretize the ideas
they encounter in lectures and assignments.  Handwritten formulas and
proof steps do not give students immediate correctness feedback.  This
can slow their development of confidence in formal reasoning and makes
assessment harder to apply consistently at scale.  Logicat addresses
this by allowing students to manipulate formulas, transformations,
clauses, valuations, and proof steps as computational objects in Racket,
building directly on their Scheme/Racket experience from the first-year
curriculum.
The system is lighter than a general proof assistant such as Coq or Lean and
uses notation aligned with the course.  It exposes composable functions
students can invoke individually or use to program their own automations.

The paper presents the design rationale, system organization, and planned
course integration of Logicat as a practical model for using
computational tools to support engagement, conceptual concreteness, and
more consistent assessment in large formal-reasoning courses.
\end{abstract}

\spacetune{\vspace{2\baselineskip}}

\spacetune{\vspace{2\baselineskip}}
\section{Introduction}

Large required courses in theoretical computer science have to make
abstract material usable for students while also supporting assessment at
scale.  In a course on logic, both problems are underscored by the same
feature: correctness depends on small formal details.  
Students
often discover such errors only after graded work is returned or, worse, not at all because the errors are missed by the graders ploughing through a large volume of handwritten work.

This paper describes Logicat, a Racket toolkit developed for CS 245,
\textit{Logic and Computation}, at the University of Waterloo
\cite{logicat2026}.  CS 245 is required for undergraduate computer
science students and develops mathematical reasoning skills through
propositional logic, first-order logic, formal proof, computability, and
program verification \cite{cs245fall2025}.  The recommended text is Lu's
\textit{Mathematical Logic for Computer Science}, used mainly for
definitions, notation, and formal deduction, with additional material
developed through lectures and course notes \cite{cs245fall2025,lu1998}.
\spacetune{\enlargethispage{\baselineskip}}

The aim of Logicat is to make course objects computationally concrete.
A formula can be entered, parsed, inspected, transformed, treated under
valuations, converted to normal forms, used as a clause in a resolution
procedure or as part of a formal deduction proof.  These are the
same objects students see in lectures and assignments, but now with
operations they can apply and test.

The use of Racket is part of the design.  Waterloo computer science
students have experience with Scheme/Racket from their first-year
curriculum, and Racket is well suited to symbolic data and functional
composition \cite{kelsey1998,flatt2010}.  Logicat therefore presents
logic through a programming medium that students already know.  The tool
is provided as a set of functions that students can call, combine,
and inspect.

The system also addresses a grading problem.  In large courses, teaching
assistants must repeatedly judge whether symbolic steps are well formed,
whether transformations preserve equivalence, whether clauses have been
resolved correctly, and whether a proof line follows from the cited
rule.  This is tedious and error-prone. Logicat can check these conditions automatically.  The
instructor and teaching assistants can then spend more attention on
explanation, strategy, and conceptual understanding.

Logicat was made available on an optional basis during a recent
offering of the course and is being prepared for broader integration.
This paper gives a design and system account.  It focuses on the parts
of the toolkit that most directly support the instructional goals:
logical formula transformation, reasoning under valuations, resolution procedures and formal deduction proofs.
Other capabilities, such as simplification of normal forms, 
are treated as supporting infrastructure rather than as the main emphasis.

\newpage
\section{Course Context and Instructional Problem}

Waterloo's CS 245 introduces logic as a tool for representation, reasoning, and
computation.  The official course description emphasizes mathematical
reasoning, propositional and first-order logic, the distinction between
syntax and semantics, formal proof, and the limits of computation
\cite{cs245fall2025}.  The course outline includes logical connectives,
truth tables, well-formed formulas, semantics, logical equivalences,
normal forms, Boolean algebra, formal deduction, soundness and
completeness, resolution, the Davis--Putnam procedure, first-order
logic, Peano arithmetic, undecidability, and Hoare-style program
verification \cite{cs245fall2025}.

This range of topics asks students to move among several views of the
same formal material:  A formula has a syntactic structure.  It has a
truth value under a valuation or interpretation.  It can be transformed
by equivalence laws.  It can be converted to a normal form and used as
input to an algorithm.  It can appear as a premise or conclusion in a
formal proof.  Students learn the topics separately, but the course
expects them to see how the views fit together.

Handwritten work makes this integration difficult.  A student may write
several lines after an early invalid transformation, or may build a
formal proof whose third line is not actually justified by the first two.
The feedback may arrive days later, if at all.  At that point the student has lost
the connection between the error and the reasoning decision that
produced it.  A small formal error becomes a delayed and often opaque
comment in the margin.

The same issue appears in assessment.  A large course requires a large
grading team.  Even with clear rubrics, judgments are hard to apply uniformly.  One grader may treat a variation as harmless
where another sees an ambiguity.  One may give generous partial credit for a
proof with an invalid early step, while another may treat the later lines
more strictly.  More importantly, grading handwritten logic under time pressure is error-prone. These aspects are not failures of goodwill --- they
are consequences of requiring a team to grade dense logic work by hand.

Logicat was designed to reduce the delay and variability of such evaluation.
Students still write proofs and explanations, but they also have
a way to test the formal objects that their explanations refer to.  In a
large class, this can change the students' learning process and graders' workflow and accuracy.  Some errors can be
found while students are still thinking, and graders can check results more easily, or even have them pre-screened. 

\section{Pedagogical Aims}

\subsection{Concretizing Formal Objects}

Students first meet logical formulas as strings of notation.  To
reason with them reliably, they must see more structure.  A formula has an inductive structure.
A rule application has a scope.  A resolution step has two
parent clauses and a resolvent.  A proof step has cited premises, a rule,
a substitution, and a conclusion.  Logicat makes such structure
accessible through data objects and operations.

This matters because many errors in beginning logic are structural.  A
student may apply De Morgan's law to wrong subformulas, distribute
over the wrong connectives, mis-parenthesize, cite a rule whose
requirements have not been met, or resolve two clauses on a symbol that
does not occur with opposite signs.  Immediate checks on such steps help
students separate two questions: whether they chose a good overall
strategy, and whether a particular operation was legitimate.

\subsection{Building on Racket Experience}

Logicat is a Racket library rather than a standalone application.  This
choice keeps the tool close to the programming experience students
already have.  A student can enter a formula as a string, inspect its
internal representation, call a transformation function, or write a
small program that performs a sequence of desired steps.

The design also lets the course use programming as a way to understand
logic.  A formula can be traversed and transformed.  A resolution
procedure can be read as a program over clauses.  A proof checker can be
connected to the same rule-instantiation discipline students are expected
to follow by hand.  The point is not to replace formal reasoning by
programming, but to let students use a familiar computational environment to
work with the formal objects of the course.

\subsection{Supporting Assessment at Scale}

Logicat can provide local correctness checks before grades are assigned.
It can check whether formulas are well formed, whether transformations are valid,
whether two propositional formulas agree under valuations, whether a
resolution step is valid, or whether a formal deduction step follows
from its justification.  These checks leave the quality of a student's
explanation for human judgment, while removing uncertainty about many formal details.

This gives the grading process a better starting point and overall reliability.  Students can
submit work that has already passed some mechanical checks, and graders
can focus on the remaining questions: the choice of method, the clarity
of explanation, and the student's understanding of the reasoning.

\section{System Organization}

Logicat represents logical formulas as structured Racket data.  Atomic
propositions are represented by symbols; compound formulas are
represented by expressions whose first component is an operator and
whose remaining components are arguments.  For example, the formula
\[
        A \rightarrow (B \land C)
\]
has the internal representation
\[
        \texttt{(impl A (and B C))}.
\]
The representation is simple enough for students to inspect directly and
regular enough to support parsing, formatting, traversal,
transformation, evaluation, and proof checking.

The system accepts input as text strings or as Racket data structures.  Students may
write formulas using Unicode logical symbols, ASCII alternatives, or
LaTeX syntax.  This flexibility is useful because logic texts and software use a wide range of notation \cite{lu1998,huth2004,velleman2019}.
Logicat normalizes inputs to a common internal
form, while output routines can print formulas in a logical notation,
Boolean-algebra notation, or LaTeX.
This allows students to copy inputs and outputs into assignments and notes.  

The supporting operations include formula input and output, structural
inspection, truth valuations, truth tables, normal-form conversion,
application of specific equivalence rules,
and automatic simplification routines.
Propositional logic is supported most fully, while formal deduction for first-order logic is supported as needed for the course.

Additionally, an interactive logical formula editor allows students to apply transformations to whole formulas and sub-expressions as needed.

These features are useful
in assignments, but they are mainly infrastructure for the parts of the
system where students work with course reasoning: local transformations,
resolution, and formal deduction.  
The rest of the paper concentrates on
those parts.

\section{Representations Used by the Reasoning Tools}

The choice of data representation is kept visible because it helps students
connect the course topics.  
A formula entered in course notation is not
stored as a string.  It is stored as a structured expression whose
operator and arguments can be extracted.  This is a basic implementation point, but it also has a pedagogical value:  
it keeps students focused on formula structure.

The representation also uses $n$-ary forms for associative connectives
such as conjunction and disjunction.  This avoids forcing students to
choose an artificial binary parenthesization when the course treats a formula
such as \(A \land B \land C\) as a conjunction of three parts (after associativity is established). 
At the
same time, operations are available to re-associate a formula when a
binary rule or a particular proof format requires it.  Thus the system
can respect the informal classroom convention while still exposing the
formal structure.

For resolution and the Davis--Putnam procedure in propositional logic, Logicat uses a separate
term representation for conjunctions and disjunctions of literals. 
A term records, for each propositional symbol, whether the symbol appears
positively, negatively, or not at all.  This is a compact way to
represent clauses and implicants.  It also makes resolution easier to
explain computationally: two clauses resolve on a symbol when that
symbol appears positively in one clause and negatively in the other, and
the resolvent is obtained by combining the remaining literals, unless a
tautology is produced.

This representation is useful in teaching because it gives students a
concrete view of clauses where the set notation can be confusing.
A clause can
be displayed as a formula, but it can also be displayed compactly as a row of signs
relative to a fixed symbol order.  The latter form is close to the way
an algorithm processes the data and makes algorithm traces intelligible.  Students can therefore compare the
mathematical rule with the data structure that implements it.

Formal deduction uses another family of objects.  Entailments, rules,
substitutions, justifications, steps, and proofs are represented
explicitly.  These objects correspond closely to the annotations students
write in a proof: the line number, the statement proved on that line, the
rule used, the cited previous lines, and the instantiation of the rule.
By making these parts explicit, Logicat can check validity without
requiring any prose explanation.

These representation choices support the same instructional pattern.
Students begin with familiar notation, but the tool enforces the structure
that notation abbreviates.  Once that structure is visible, the student
can ask more precise questions: which subformula is selected, which
clauses are being resolved, which substitution instantiates the rule, or
which earlier lines justify the present one.  Those questions are 
where students often make errors.

\begin{figure*}
\begin{verbatim}
;; Apply the Davis-Putnam procedure on the formulas in lf-list,
;; eliminating proposition symbols in  the order given by sym-list.
(define (dpp atom-list lf-list)
   (define (make-clause lf) (make-term-with-lf 'or atom-list lf))
   (define ((has-pos? ix) term) (equal? tlit-pos (term-ref term ix)))
   (define ((has-neg? ix) term) (equal? tlit-neg (term-ref term ix)))

   (let* ([cnf-list   (map lf-cnf lf-list)]
          [conjuncts  (map lf-conjuncts cnf-list)]
          [uconjuncts (apply lset-union #:equal? lf=? conjuncts)]
          [clauses    (filter term? (map make-clause uconjuncts))])
     (do ([i 0 (+ 1 i)])
       [(= i (length atom-list))]
       (let* ([atom (list-ref atom-list i)]
              [pos-atom-clauses (filter (has-pos? i) clauses)]
              [neg-atom-clauses (filter (has-neg? i) clauses)]
              [atom-clauses (append pos-atom-clauses neg-atom-clauses)]
              [resolvents '()])
         (for ([pos pos-atom-clauses])
           (for ([neg neg-atom-clauses])
             (let ([r (resolve-terms i pos neg)])
               (when (term? r) ;; Exclude #t terms.
                 (set! resolvents (lset-union1 r resolvents)) )) ))
         (set! clauses (lset-minus clauses atom-clauses))
         (set! clauses (lset-union clauses resolvents  )) ) )
     ;; At this point clauses is either empty
     ;; or has only the empty clause.
     (if (null? clauses) clauses (list (lf-or))) ))
\end{verbatim}
\caption{Programming example: the Davis--Putnam procedure}
\label{fig:dpp}
\end{figure*}

\section{Formula Transformations and Feedback}

Formula transformation is the first place where Logicat gives students
immediate feedback on course-level reasoning.  The system provides
transformation rules for a range of properties proven in the course: implication elimination, equivalence elimination,
double-negation removal, associativity, commutativity, distribution, and
De Morgan transformations.  Some rules act at the top level of a formula;
others operate recursively, for example to eliminate non-basic
connectives, push negations down to atoms, or convert formulas to DNF or
CNF.

The local nature of rule application is important.  In handwritten work,
students may write a transformed formula without recording exactly which
subformula changed.  Logicat makes the selected subformula explicit.  A
student can select a part of the formula, apply a rule to that part, and
see the new whole formula.  This gives a concrete counterpart to the
usual classroom instruction that equivalence transformations must be
applied to well-defined subexpressions.

Logicat also includes a small formula editor.  The editor maintains a
current formula and a current selection.  Students can move the selection
within the formula tree, grow it across adjacent arguments of an
associative connective, and apply a rule to the selected part.  

For propositional logic, truth tables and normal forms remain available as checks.  A student can
transform a formula and then test whether the result is equivalent to
the original.  The tool can also show valuations under which two formulas
differ.  These checks are simple, but they often reveal the kind of 
error that would otherwise be found only in high-quality grading.

This use of the tool also helps move students toward planning.  If local
steps are checked as they are made, students can spend less effort
wondering whether a copied connective or a local equivalence would  be valid,
and more effort deciding what form they are trying to reach.  In a
normal-form exercise, for example, the question becomes not only ``what
is the next legal rewrite?'' but ``which legal rewrite moves the formula
toward the intended shape?''

\section{Programmability}

Students can write their own Racket programs to perform sequences of steps.   This can be useful when they wish to put formulas into some particular form, or for more ambitious purposes.

They can also inspect example programs that embody the algorithms taught in the course.
One such example is the Davis--Putnam procedure for resolution theorem proving.   We use this as an illustration.

The \texttt{dpp} procedure converts formulas to CNF, extracts clauses, partitions clauses
by positive and negative occurrences of a symbol, resolves pairs, and
adds the resolvents back to the clause set.
Figure~\ref{fig:dpp}
shows the structure of the
implementation.  Details such as tracing and error handling are omitted, as are the definitions of the subordinate functions.
An important point is that the code uses the same exposed operations that are available elsewhere in the library, \textit{e.g.} \texttt{lf-cnf} which converts a logical formula to conjunctive normal form.

The example is included because it shows the role of composability.  CNF
conversion, conjunct extraction, term construction, term inspection, set
operations on lists, and resolution are separate pieces.  The
Davis--Putnam procedure is obtained by composing them.  This makes the
algorithm available at two levels.  A student can run it as a command to
check an example, or can read and modify the program to understand how
resolution is being applied.

When run in verbose mode (the code for which is not shown), the procedure displays formulas, clauses,
selected symbols, resolving pairs, resolvents, and the remaining clause
set.  This trace is useful in a course setting because it connects the
formal resolution rule to the changing state of the algorithm.  The
trace also helps students see why an implementation must make choices
that are easy to omit in a paper proof, such as the order in which
symbols are eliminated and the treatment of tautological resolvents.

The same code can also become an assignment object.  Students can be
asked to run the procedure on a small inconsistent set, explain the
empty-clause result, change the symbol order, or compare the trace with a
manual resolution derivation.  These are modest programming tasks, but
they require students to connect the course rule to an executable
procedure.

\begin{figure*}
\begin{minipage}[t]{.495\textwidth}

Successful proof: substitutions inferred
\begin{verbatim}
(fd-check-proof (mk-fd-proof '(
  [1 "A->B, B->C, A |- A->B"
     (mem)      ]
  [2 "A->B, B->C, A |- A"    (mem)      ]
  [3 "A->B, B->C, A |- B"    (impl- 1 2)]
  [4 "A->B, B->C, A |- B->C" (mem)      ]
  [5 "A->B, B->C, A |- C"    (impl- 4 3)]
  [6 "A->B, B->C |- A->C"    (impl+ 5)  ]
)))
\end{verbatim}

Output:
\begin{verbatim}
Checking step 1: Subst found
  A = A->B, Sigma = {B->C,A}. OK
Checking step 2: Subst found
  A = A, Sigma = {A->B,B->C}. OK
Checking step 3: Subst found
  A = A, B = B,
  Sigma = {A->B,B->C,A}. OK
Checking step 4: Subst found
  A = B->C, Sigma = {A->B,A}. OK
Checking step 5: Subst found
  A = B, B = C,
  Sigma = {A->B,B->C,A}. OK
Checking step 6: Subst found
  B = C, A = A,
  Sigma = {A->B,B->C}. OK
Proof OK!
\end{verbatim}
\end{minipage}
\begin{minipage}[t]{.495\textwidth}
Failing proof: explicit substitutions
\begin{verbatim}
(fd-check-proof (mk-fd-proof '(
  [S1 "A->B, B->C, A |- A->B" (mem)
      "A=A->B, Sigma={B->C,A}"       ]
  [S2 "A->B, B->C, A |- B"    (mem)
      "A=A, Sigma={A->B,B->C}"       ]
  [S3 "A->B, B->C, A |- B"    (impl- S1 S2)
      "A=A, B=B, Sigma={A->B,B->C,A}"]
  [S4 "A->B, B->C, A |- B->C" (mem)
      "A=B->C, Sigma={A->B,A}"       ]
  [S5 "A->B, B->C, A |- C"    (impl- S4 S3)
      "A=B, B=C, Sigma={A->B,B->C,A}"]
  [S6 "A->B, B->C |- A->C"    (impl+ S5)
      "A=A, B=C, Sigma={A->B,B->C}"  ]
)))
\end{verbatim}

Output:
\begin{verbatim}
Checking step S1: Subst given. OK
Checking step S2: Subst given. Failed
Proof fails at step S2.
Proof failed.






\end{verbatim}
\end{minipage}

\caption{Formal deduction examples of correct and incorrect proofs}
\label{fig:fd-proofs}
\end{figure*}

\section{Formal Deduction Checking}

Formal deduction is the part of the course where delayed feedback can be
most damaging.  Students often write incorrect proofs with plausible overall
shape, but where an early line does not follow from its cited rule.  Several
later lines may then depend on a statement that has not been established.  
Delayed feedback means that students do not get sufficient practice at discriminating between valid and invalid operations.
One of the principal motivations in the development of Logicat was to help students learn to navigate this challenge.

Logicat represents the elements of a formal deduction proof explicitly:
entailments, rules, substitutions, justifications, steps, and proofs.  An
entailment has the form
\[
        P_1,\ldots,P_n \vdash C,
\]
where the premises may be formulas or symbols representing sets of
formulas, and the conclusion is a formula.  A rule specifies required
entailments and a resulting entailment.  A proof step contains a claimed
entailment, a justification naming a rule and previous steps, and a
substitution showing how the rule is instantiated.

The checker verifies proof steps in sequence.  It checks that the cited
rule exists, that the cited previous steps exist, and that the rule's
requirements and result match the cited entailments and the current
claim after substitution and canonicalization.  If the substitution is
not supplied, the system infers it if possible.  The result is a
check on whether a proof step has been justified.

Figure~\ref{fig:fd-proofs} gives examples of correct and incorrect proofs of the transitivity of implication in the
course's formal-deduction notation.  The first proof uses the theorem
\texttt{mem} for membership in the premise set, implication elimination
\texttt{impl-} $(\rightarrow-)$ and implication introduction \texttt{impl+} $(\rightarrow+)$.
The student supplies the proof steps but not the substitutions; Logicat
finds the rule instantiations.

The checker returns \texttt{\#t}.  In verbose mode it also reports the
substitutions that it inferred for the rule applications, as shown.
This output is useful pedagogically because it shows more than success or
failure.  Students can verify how a schematic inference rule is matched to
a concrete line of proof.  The substitutions make explicit what is often
left implicit, and often wrong, in handwritten work.  The set of surrounding
premises, represented here by \texttt{Sigma}, is part of the rule
instantiation rather than a background assumption.
Note the ASCII forms ``\texttt{->}'' and ``\texttt{|-}'' in the example are accepted input forms for
the course notations ``$\rightarrow$'' and ``$\vdash$''.

The system also allows substitutions to be supplied explicitly.  This is
useful for debugging, for tests, and for assignments in which the
instantiation itself is part of what the student is expected to provide.

The second ``proof'' has explicit substitutions, but line
\texttt{S2} incorrectly claims membership of \(B\) from premises that do
not include \(B\).
The checker returns \texttt{\#f} and reports the first failing line.

This example illustrates the assessment problem at the scale of the
course.  A short proof of six lines is easy to inspect after the error is
pointed out.  In assignments, proofs may be twenty or thirty lines, with
later lines depending on earlier ones.  A graduate teaching assistant can
find such errors, but doing so reliably across hundreds of submissions is
slow and uneven.  The checker gives a uniform answer to the local
question: does this line follow from the cited rule and previous steps?

It also changes the student's task.  Without checking, students often
produce a sequence of lines and hope that the derivation will be accepted
as a proof.  With checking, they can find  and correct their mistakes while working.
The work shifts toward proof planning: identifying useful intermediate
claims, choosing when to introduce or eliminate a connective, and
arranging the proof so that it follows from justified steps.
The checker does not choose the plan, but it prevents unsupported lines
from silently becoming part of the plan.

This differs in scope from using a full proof assistant.  Systems such
as Coq and Lean support machine-checked mathematics and verified
software at a much larger scale \cite{barras1997,demoura2015}.  They
also bring their own syntax, libraries, tactic languages, and proof
cultures --- and consequently would take considerable course time to introduce.
Logicat is designed for the notation and inference rules of
the specific course.  It gives students feedback on the formal steps used in the
course without making the operation of an advanced proof assistant itself a major topic.

\section{Using Checked Work in Learning Activities}

The examples above suggest several ways to use Logicat in assignments
without changing the character of the course.  The tool can be used
privately by students while they work, as a source of transcripts or
program fragments in submitted work, or as a checking aid for graders.
These uses can be combined gradually.  A first assignment might ask
students only to use the tool to confirm a transformation.  A later
assignment might ask them to submit a short Racket expression that builds
and checks part of a proof.

One useful pattern is a guided transformation exercise.  Students are
given a formula and a target form, such as a formula using only
\(\neg\), \(\land\), and \(\lor\), or a formula in conjunctive normal
form.  They must record the transformations they chose, but they can use
Logicat to check each local application.  The written part of the
answer can then ask why those transformations were chosen.  This keeps
the emphasis on reasoning while reducing time spent on uninformative
copying errors.

A second pattern is a resolution exercise.  Students can first perform a
manual resolution derivation and then compare it with a Davis--Putnam
trace.  If the two differ, the question becomes diagnostic: did the
manual proof use a different elimination order, did it omit a resolvent,
or did it introduce a clause that does not follow?  The tool does not
replace the manual derivation; it gives students another representation
of the same process.  This is particularly helpful because resolution is
both a proof rule and an algorithmic procedure.

A third pattern is proof repair.  Students can be given a nearly correct
formal deduction proof containing one or two invalid lines.  Their task
is to run the checker, identify where the proof first fails, explain why
the cited rule does not justify the line, and repair the proof.  This
activity is close to what teaching assistants do when grading.  It also
helps students see that a proof is not just a sequence of plausible
statements.  Each line has a local contract: a rule, cited previous
steps, and an instantiation.

A fourth pattern uses the programmability of the library.  Students can
write small Racket functions that generate families of examples, apply a
normalization procedure, or test a conjectured equivalence over a list of
formulas.  Such exercises use the students' existing programming
background to deepen their contact with the formal material.  This is
consistent with evidence that active engagement improves learning in
STEM courses \cite{prince2004,freeman2014}, and with previous work using
interactive computational tools in theoretical computer science topics
\cite{cavalcante2004}.

These activities have a common form.  The student is not asked simply
to produce a final formula or a final proof.  The student is asked to
work with intermediate objects: selected subformulas, clauses,
resolvents, substitutions, and checked proof lines.  Those intermediate
objects are where much of the learning happens.  They also give the
course staff better evidence about what the student understood and where
an error entered the reasoning.

For a large required course, this matters.  Some students will quickly
see how to extend the library, while others will only use the examples
provided with the assignments.  Both levels of use are useful.  The
basic commands give immediate feedback on local correctness.  The
programming interface gives stronger students a way to explore the logic
more deeply, while students who are less enthusiastic about the course still benefit from more engagement.

\section{Course Integration and Assessment Workflow}

Logicat was introduced on an optional basis in a recent course
offering and is being prepared for broader integration.  The initial
use was deliberately conservative.  Students could use the system to
explore examples and check work, but success in the course did not
depend on adopting a new software workflow before the documentation
and assignment design had stabilized.

The planned integration is staged.  Early use should be light: students
enter formulas, run simple checks, and compare the output with what they
have done by hand.  Later assignments can require more structured
artifacts, such as a transcript of transformations, a checked proof
fragment, or a short Racket expression that constructs and verifies part
of a solution.  This progression allows students to begin with direct
commands and move gradually toward more programmatic use.

This staged approach is important in a required course.  Some students
will quickly use the library as a programming environment and write
their own helpers.  Others will use only the commands needed for an
assignment.  Both levels of use are acceptable.  The basic commands give
immediate feedback on local correctness, while the programming interface
gives stronger or more curious students a way to explore the formal
material more deeply.

The submitted work can still include handwritten or typeset
explanations.  The difference is that some formal details can be checked
before submission.  A student building a formal deduction proof, for
example, can verify that each line follows from its cited rule and
previous lines while the proof is being written.  If a step fails, the
student receives feedback at the point where the reasoning decision was
made, rather than days later in a grading comment.

The assessment workflow can use the same artifacts at several levels.
First, students may use Logicat privately as a checking aid.  Second,
an assignment may ask for selected output or code showing how part of a
solution was checked.  Third, course staff may re-run selected checks
when grading.  In each case, the tool supplies a consistent answer to
local formal questions, while teaching assistants remain responsible for
evaluating explanation, strategy, and understanding.

This changes the role of grading without removing human judgment.  The
teaching assistants need not be the first source of feedback on every
syntax error, invalid transformation, or unjustified proof line.  Nor
need they rely only on visual inspection of a long proof that may contain
an early unsupported step.  Their effort can be directed toward the
questions that require judgment: whether the student chose an appropriate
method, explained the reasoning clearly, and reached the required result
in a principled way.

\section{Limitations and Future Work}

Logicat is still under development.  Its mature components support
propositional formula manipulation, local transformations, normal forms,
simplification, resolution, Davis--Putnam-style consistency checking,
and formal deduction checking in the style used by the course.  First-order
formal deduction and resolution proving has recently been added, as well
as program verification as used in Waterloo's CS 245.

The system also requires careful assignment design.  Running commands is
not evidence of understanding by itself.  Assignments must ask for
explanations, intermediate choices, and interpretation of results, with
Logicat used to check and explore the formal objects involved.  This is
especially important for formal deduction: a correct transcript can show
that a proof is justified, but students must still explain why
the chosen intermediate claims are useful.

A further limitation is that the current paper does not report measured
learning gains.  The next stage is full integration into the course,
followed by evaluation.  Possible measures include student engagement,
patterns of tool use, reduction in common local errors, grading
consistency across teaching assistants, and student attitudes toward
formal reasoning.  Active-learning studies in STEM education provide a
useful background for this evaluation, but course-specific evidence will
be needed \cite{prince2004,freeman2014}.  Work on interactive tools for theoretical computer science,
such as JFLAP, also provides a point of comparison for studying how
computational artifacts affect student work in formal subjects
\cite{cavalcante2004}.

Future development will focus on three practical issues.  The first is
making assignment templates that use the tool without turning assignments
into software tutorials.  The second is improving diagnostics so that
students receive feedback that is useful without giving away the proof
strategy.  The third is improving support for the first-order material in
the course, including richer syntax and proof patterns.

\section{Conclusion}

Logicat was developed to help students in a large required theoretical computer science course engage more actively.  Logical formulas, proofs  and related objects
are represented as data structures in Racket.  This builds on students'
first-year Scheme/Racket experience and gives them immediate checks on
many of the formal details that otherwise receive delayed feedback.

The strongest uses of the system  are 
where students see reasoning as a sequence of checkable operations:
selecting and transforming a subformula, following a resolution trace,
or building a formal deduction proof whose lines are justified.  These
activities help move student work from producing strings of symbols
toward planning and checking a derivation.

The same properties support assessment at scale.  Many routine formal
judgments can be made accurately and consistently by the tool, leaving teaching
assistants to evaluate explanation, strategy, and understanding.  Full
course integration and evaluation remain future work, but the design
provides a practical way to connect programming experience with formal
reasoning in a large logic course.

We acknowledge that parts of this manuscript were drafted with the aid of AI tools.

\balance
\begingroup
\normalsize
\bibliographystyle{IEEEtran}
\bibliography{main}
\endgroup

\end{document}